# Deep sub electron noise readout in CCD systems using digital filtering techniques.

Gustavo Cancelo[1], Juan Estrada[1], Guillermo Fernandez Moroni[2], Ken Treptow[1], Ted Zmuda[1], Tom Diehl[1].

[1] Fermi National Accelerator Laboratory, Batavia, Il 60510, USA,
Kirk Road and Pine Street, Batavia, Il 60510, USA
cancelo@fnal.gov, estrada@fnal.gov, treptow@fnal.gov, zmuda@fnal.gov, diehl@fnal.gov.

[2] Instituto de Investigaciones en Ing. Eléctrica (IIIE) "Alfredo C. Desages"
Departamento de Ingenieria Eléctrica, Universidad Nacional del Sur
Av. Alem 1253 - (B8000CPB) Bahıa Blanca, Argentina
fmoroni.guillermo@gmail.com

**Abstract.** Scientific CCDs designed in thick high resistivity silicon (Si) are excellent detectors for astronomy, high energy and nuclear physics, and instrumentation. Many applications can benefit from CCDs ultra low noise readout systems. The present work shows how sub electron noise CCD images can be achieved using digital signal processing techniques. These techniques allow readout bandwidths of up to 10 K pixels per second and keep the full CCD spatial resolution and signal dynamic range.

## 1 Introduction

CCDs were invented in 1969 (Amelio1970; Boyle 2010; Smith 2010) and today, after 4 decades of advances in R&D, they have become the sensor of choice for a large number of current and future ground and space telescopes (Flaugher 2006; Honscheid 2008; Diehl 2008). Scientific CCDs are developed with high performance characteristics such as high quantum efficiencies above 90% in the visible and infrared spectrum, large photon dynamic range exceeding 100,000 $e^-$, and pixel size around 10 μm. In particular, CCDs designed at Lawrence Berkeley National Laboratory (Holland 2003) are made in 250 μm thick high resistivity Si. These detectors work fully depleted improving the quantum efficiency in the infrared up to 1000nm. Furthermore, the 250 μm thickness makes these CCDs excellent detectors for a number of high energy and nuclear physics applications. For those applications the important features, rather than optical performance, are the CCD low noise which allow detection of low ionizing particles, the high spatial resolution, and mass. Just to mention one example, the DAMIC experiment at the Fermilab National Accelerator Laboratory searches for Dark Matter measuring the nuclear recoil energy of WIMPs with the Si atom nuclei (Estrada 2008). A sub electron noise CCD readout system allows sensitivities in the few eV region, which makes DAMIC the most sensitive detector for low mass WIMP search. Sub electron noise CCD readout systems are also important in the design of a new generation of optical spectrometers (Chandler 1990; Janesick 1990). In a spectrometer the required collecting area of the telescope is proportional to the signal to noise ratio.

The reminder of this paper is organized as follows. Section 2 introduces the CCD signal and noise problem. Section 3 shows how the correlated noise limits the CCD readout noise. Section 4 describes a new technique to lower the correlated noise. Section 5 shows the achieved sub electron noise results. Section 6 describes the errors of the method and Section 7 shows the hardware implementation of a complete system.

## 2 CCD signal and noise

CCDs are organized in one or more arrays of small pixels (e.g. 15 x 15 micron pixels). When properly biased, CCD pixels store charge in a potential well. There are several ways charge can be generated (Janesick 2001; McLean 2008) and once is collected, the charge can be readout by sequentially shifting it along the CCD vertical columns into an horizontal serial register which is clocked out through a video amplifier one pixel at a time. Some CCDs have multiple video outputs which allow splitting the array in blocks and reading them in parallel (Figure 1).

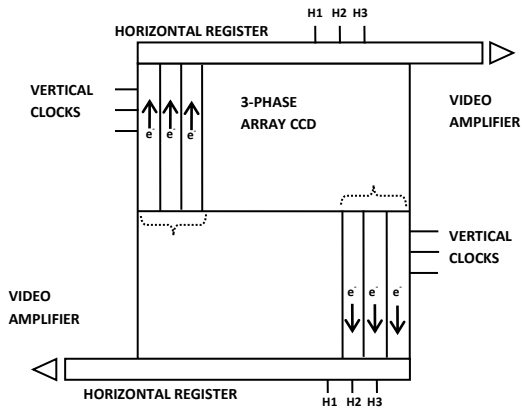

Fig. 1: CCD array showing 2 sets of video channels, vertical and horizontal clocks.

A CCD transfer function computes the analog and digital stage gains (Figure 2). At high signal levels the noise is dominated by photon statistics and pixel layout uniformities. At very low signal levels the noise is dominated by the readout electronics. The CCD gain is defined in electrons per digital units (e-/ADUs) which is the inverse of the product of the individual block gains (Figure 2).

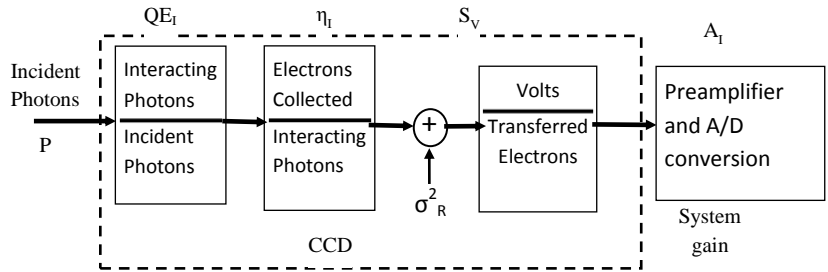

Fig. 2: CCD Block diagram and transfer functions.

When the CCDs are cooled with liquid nitrogen at temperatures of about 173°K the dark current is negligible. Furthermore, the charge transfer efficiency (CTE) from pixel to pixel is better than 99.9998% with a total noise contribution of about 0.1%. Hence, the main noise contributors to the CCD data are the internal JFET video amplifier and the external readout system.

The CCD output stage generates a video signal as shown in Figure 3. A reset pulse sets a reference level before a pixel charge is transferred to the sense node capacitor via the output summing well. The video is sensitive to power supply and output JFET noise, in particular to their low frequency content.

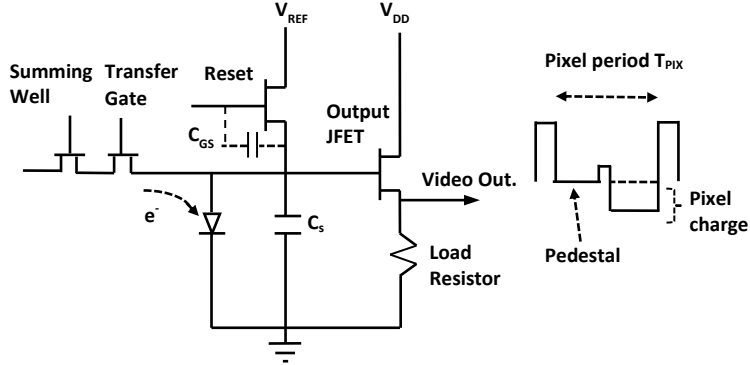

Fig. 3: Output amplifier region and video signal showing clock feed-throughs.

The most common signal readout technique is called Correlated Double Sampling (CDS) which subtracts the average pedestal from the average pixel signal during a predefined time window (Equation 1). Figure 4 shows the CDS transfer function as a filter response. CDS is particularly sensitive to low frequency content noise near $0.4/T_\int$. $T_\int$ is the CDS integration time. The CDS differential integral has zeros at frequencies multiple of $T_\int$.

$$s_j^{cds} = \int_{t_{oj}}^{t_{1j}+T_\int} p_j + n_{total}(t)\,dt - \int_{t_{2j}}^{t_{3j}+T_\int} s_j + n_{total}(t)\,dt \qquad (1)$$

Where $p_j$ and $s_j$ are the j-th pedestal and pixel respectively, $n_{total}(t)$ represents the total additive noise (correlated and uncorrelated), and $t_{oj}$, $t_{1j}$, $t_{2j}$, $t_{3j}$ define the limits of the integration windows.

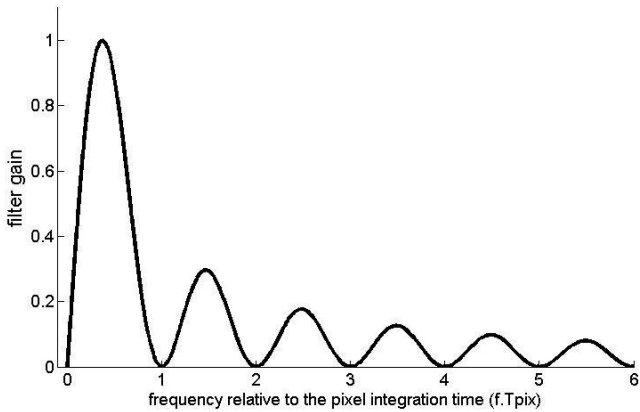

Fig. 4: CDS transfer function.

CDS is the optimal filter for a single pixel measurement. However, due to the correlated pink noise, the CDS variance does not monotonically decay towards zero as a function of the integration time ($T_j$). Indeed, the noise variance has a minimum at some finite time and then monotonically increases due to the dominant 1/f noise. The minimum noise achievable is related to the 1/f corner frequency and, using DeCam CCDs, has been measured to be between 20 and 25 μs (Estrada 2006). Since the 1/f noise is added by the CCD output amplifiers, we will show in the next sections that the 1/f noise component can be reduced by computing the correlation in a string of neighboring pixels.

## 3 Pixel correlated noise

Modern CCD readout systems such as the Monsoon for DECam (Estrada 2006) achieve about 2e- R.M.S. of noise at 20Kpix/s, but it rises considerably for slower readouts due to 1/f noise. The same system achieves about 7 e- of noise at 250Kpix/s. This noise meets comfortably the noise specifications for the DES (Flaugher 2006).

In order to achieve lower noise numbers for the applications mentioned above, it is important to attack the problem in several fronts including: the white Gaussian noise (WGN) contributions of the upstream sections of the readout system, the EMI contamination due to circuit resonances at high and low frequencies, and the 1/f noise effect on the CDS. The later is the main subject of this paper and is achieved by studying the correlation of low frequency content noise in a string of pixels.

To lower the WGN and EMI we have designed a special flex cable with balanced video circuits to minimize inductive coupling. Capacitive coupling has also been improved by location and shielding of the video signals. A dV/dt of 250 V/μs produces negligible crosstalk in the video signals.

Power supply noise compensation circuits have been applied to the most sensitive CCD analog powers (i.e. $V_{DD}$ and $V_R$).

The traditional analog CDS integrator has been replaced by a sampled system. This system employs a Σ-Δ 24 bit A to D converter sampling at 20Ms/s (Analog Devices 2006). The A/D includes internal digital filters. We used 2 FIR banks as a low pass filter (LPF) with a decimation ratio of 8, setting the Nyquist frequency at 1.25MHz. The intrinsic noise of the 24 bit A/D system is 1.5e-/sample and contributes less than 0.3e- of noise when operated in CDS at 100KHz (Figure 5). If the CCD video is pre-amplified, the A/D noise is further reduced by the pre-amplification factor. In our system a pre-amplification of 4 to 6 is used.

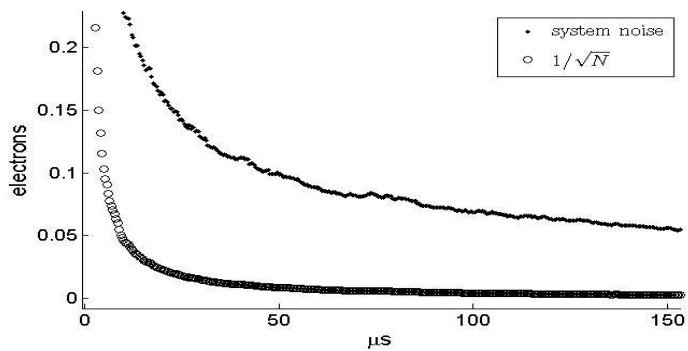

Fig. 5: Σ-Δ 24 bit A/D based readout system noise.

Figure 6 shows the noise spectrum of the digital readout system. This spectrum shows substantially lower WGN noise than a typical Monsoon system. However, despite efforts in lowering EMI resonances, we can still see 60 Hz and harmonics in the low frequency and other resonances in the higher frequency part. It is worth to mention that the high frequency noise content has a small contribution to the noise when the CCD is readout at low speed.

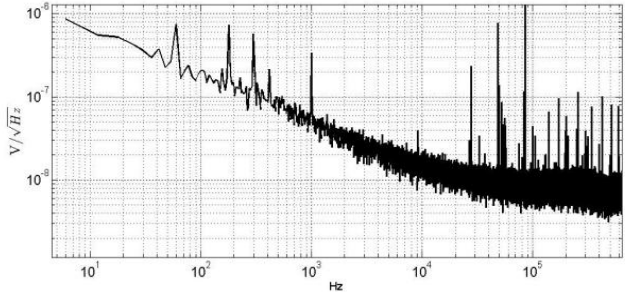

Fig. 6: Readout system noise spectrum.

An example of the video signal the Σ-Δ 24 bit A/D system samples is shown in Figure 7. Since the video data is band limited by an anti aliasing filter and fast sampled at 20Ms/s the digital CDS nicely approaches the analog CDS filtering function for the 100KHz band of interest.

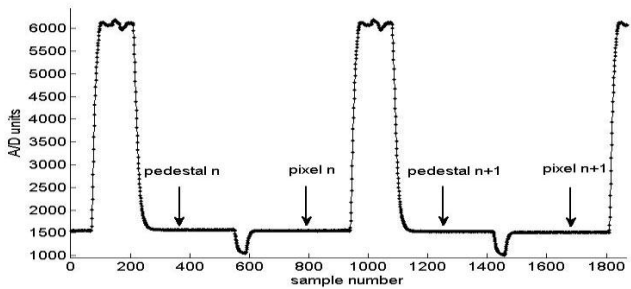

Fig. 7: Digitized video output.

The digital CDS as a function of the pixel integration time $T_f$ is shown in Figure 8. The system gain has been measured to be 0.025e-/ADU which shows an increased sensitivity due to the 24 bit ADC. The system's gain has been calibrated using a $^{55}$Fe source.

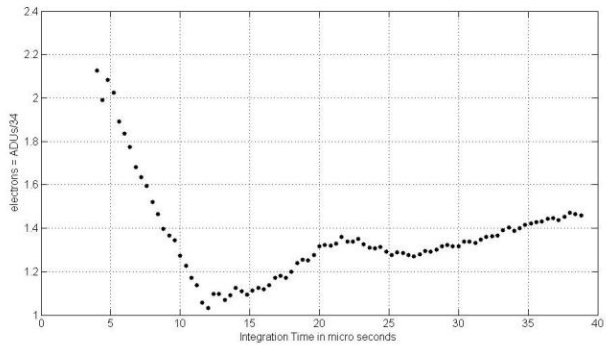

Fig. 8: Digital CDS noise as a function of the CDS integration time.

The minimum noise achieved is 1.1e- at 12µs of $T_f$. Longer integration times show the effect of the 1/f noise. DES CCDs readout times of 4 µs find noise levels approaching 2.2e-. These numbers represent a reduction of about 3 times with respect to current readout systems

## 4 Estimation of correlated noise

An additional for reducing noise was aimed at both estimating and filtering the low frequency noise content. The low frequency noise correlates among many pixels schematically shown in Figure 9. The following algorithm is used to reduce 1/f or correlated noise:

1) Digitally sample the video signal.
2) Estimate the low frequency correlated (LFC) noise of a string of pixels.
3) Subtract the correlated noise from the original video
4) Perform the digital CDS of the filtered signal.

Since the video is sampled the low frequency noise in some number of frequency bands $\Delta f = f_s/N$ can be estimated and subtracted. Thus $f_s$ is the sampling frequency and N the number of samples in the data set. For the estimation problem we use a linear $\chi^2$ estimator because it does not assume a particular noise model. $\chi^2$ estimators are not optimal but since the noise spectrum may not be completely Gaussian other estimators cannot claim optimality either.

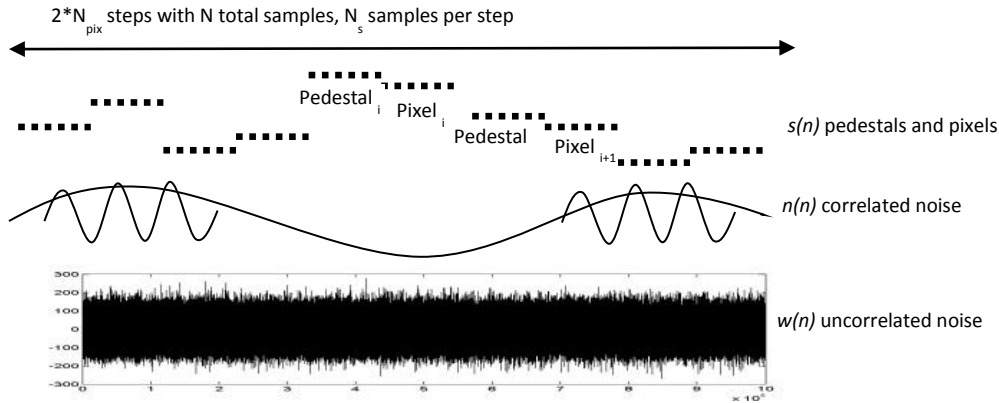

Fig. 9: Signal and noise model showing: s(n) alternated pedestals and pixels, n(n) correlated noise, and w(n) uncorrelated noise.

$N_{pix}$, $N_s$ and N are the number of pixels, the number of samples per pixel and the total number of samples used for each correlation analysis.

The signal model is described by

$$x(n) = s(n) + n(n) + w(n) \qquad (2)$$

Where s(n) is the noise free video signal composed of pedestals and pixels, n(n) is the correlated noise, and w(n) is the uncorrelated noise.
Since pedestal and pixel values are valid during a step function of uniform time we define that as:

$$U_i(n) = U(n - iN_{pix}) - U(n - (i-1)N_{pix})$$

Where $U(n-n_0)$ is the Heaviside function centered at time $n_0$.
The noise free signal can be represented by:

$$s(n) = \sum_{i=1}^{2N_{pix}} s_i U_i(n) \quad (3)$$

In equation 3 $s_i$ alternatively indexes over pedestal $i$ and pixel $i$ from $i=1$ to $2N_{pix}$.

For the correlated noise we use a Fourier base representation limited to the number of frequency modes M to be estimated:

$$n(n) = \sum_{k=1}^{M} c_k e^{2\pi f_k n} = \sum_{k=1}^{M} a_k \cos 2\pi f_k n + b_k \sin 2\pi f_k n$$

The number of parameters we need to estimate equals $p$ pixels, $p$ pedestals, and M complex frequencies or 2M real frequencies (i.e. 2p+2M total). In order to reduce the number of parameters and to improve the estimator covariance we instead use the following input signal

$$y(n) = x(n) - \langle x(n) \rangle_{U_i} \quad (4)$$

Where $\langle x(n) \rangle_{U_i}$ is the average of the signal inside each step $U_i(n)$. Hence,

$$y(n) = n(n) - \langle n(n) \rangle_{U_i} + w(n) \quad (5)$$

The correlated noise averaged in each pixel is expressed by:

$$\langle n(n) \rangle_{U_i} = \sum_{i=1}^{2N_{pix}} \left[ \left( \frac{1}{N_s} \sum_{n \in U_i} n(n) \right) U_i(n) \right] \quad (6)$$

We see that equation (6) is independent of the true pedestal and pixel values, which are accounted for separately.

The estimator model becomes:

$$y = H \cdot \theta + w \quad (7)$$

where y is Nx1 dimensional, θ is the 2M dimensional parameter vector and H is an Nx2M matrix. The H matrix is generated by H=F-E where F is a Fourier base of sines and cosines and E is a matrix of step functions whose values equal to the average of Fourier modes in each pixel interval. As a consequence H is not orthogonal. The parameter estimation is obtained by

$$\hat{\theta} = (H^T H)^{-1} H^T y \quad (8)$$

Although H is a long matrix in the N dimension the pseudoinverse has to be computed only once.
The estimation process is bounded by computing resources and numerical problems related to matrix size. We have obtained excellent results using discrete data sets of 20 pixels and 20 pedestals.

An important step in the estimation process is to determine the number of modes to be estimated. Since the filtered signal is later integrated by the CDS, the noise components are weighted by the CDS transfer function (Figure 4). Figure 10a shows a typical spectral analysis of the correlated noise for 20 pixels weighted by the CDS transfer function. Figure 10b shows the cumulative noise versus the number of spectral modes. The first 200 low frequency modes, i.e. a 20KHz total noise bandwidth, account for 85% of the low frequency noise. It is worth mentioning that these plots do not take into account errors in the estimation process which will be analyzed in Section 6.

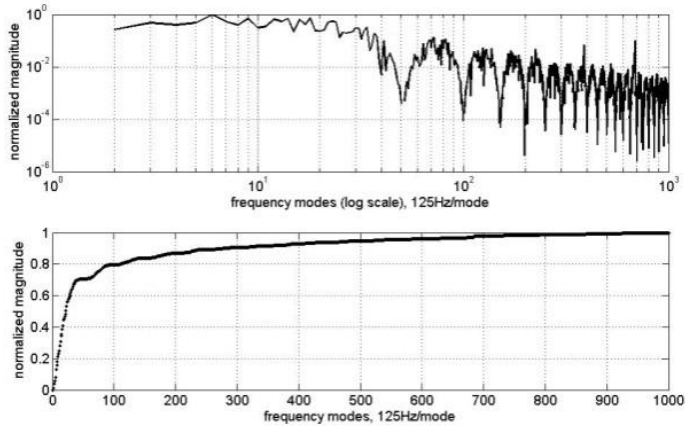

Fig. 10: a) Noise spectrum weighted by the CDS transfer function. b) Cumulative correlated noise as a function of the number of modes.

## 5 Estimation results

We have applied the estimation and filtering algorithm described in Section 4 to numerous sets of video data.
Some data sets acquired CCD and readout system noise in the video while the CCD clocks are not operating or are operated in reverse. The CCD operated in reverse give us the lowest noise values. We still get all the video amplifier, power supply noise and clock crosstalk but we do not get any charge noise due to leftover charge in the CCD. We have also applied the same methods to normal CCD readout datasets. Figure 11 shows the RMS noise of a 2 channel DES CCD operated normally as a function of the CDS window integration time. The dotted trace video has previously processed by the estimator. The trace with circles only computes the digital CDS. The processed signal achieves a minimum noise of 0.5e- at about 70μs. This is an improvement of 3 to 4 over the digital CDS.

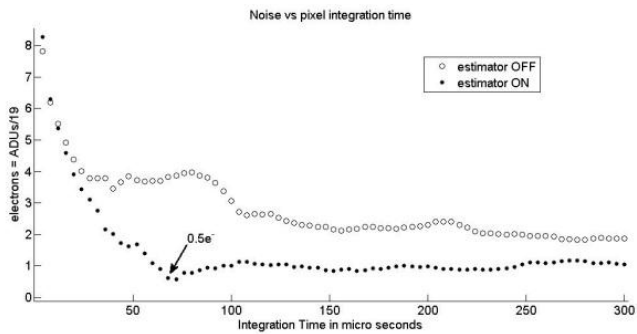

Fig. 11: Comparison of the R.M.S. noise as a function of the CDS integration time for a signal processed by the estimator and the same signal not processed by the estimator.

The picture in Figure 12 corresponds to a 12 video channel CCD designed by Berkeley National Laboratory. In this case the video data does not have the CCD clocks running but still measures all the CCD video amplifier noise and system noise, including the 1/f noise. The plot displays the average noise of 100 data sets and 1-σ error bars. The signal processed by the estimator has a minimum noise of 0.4e- at 120 μs, which is a factor of 3.5 times better than the digital CDS alone. It is also interesting that the 1-σ error bars of the estimator processed data are 4 times smaller than the ones for the unprocessed data. The plot also shows increasing noise at longer integration times due to residual 1/f noise.

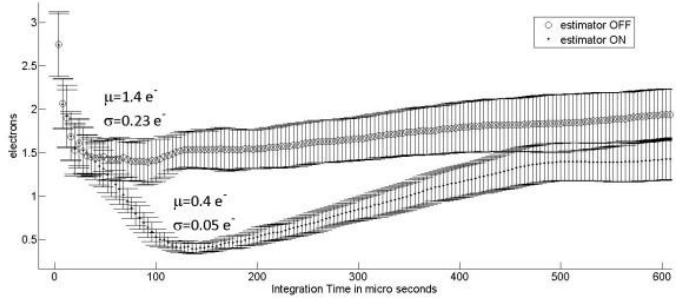

Fig. 12: Comparison of the R.M.S. noise as a function of the CDS integration time for a signal processed by the estimator and the same signal not processed by the estimator. Plot also shows 1- σ error bars.

Figure 13 compares the noise power spectrum of the unfiltered signal and the filtered signal after the low frequency estimation of 200 modes has been subtracted. On average, the LFC noise has been reduced by almost an order of magnitude on average. It can also be observed that the filter works more effectively for the higher estimated modes due to non-uniform estimation errors. Please note that the noise and spectrum plots of Figures 11 to 13 are affected by estimation errors, which are the main limitation to achieving further noise reduction.

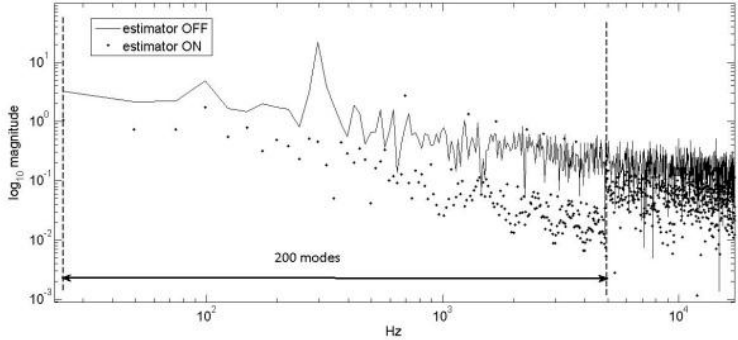

Fig. 13: Noise spectrum comparison between a signal processed by the estimator and the same signal not processed by the estimator.

## 6 Estimation error

The estimation errors can be computed using the $\chi^2$ covariance matrix
$$C_\theta = \left(H^T C_y^{-1} H\right)^{-1} \quad (9)$$
If the uncorrelated noise is WGN~$(0,\sigma^2)$ (9) can be expressed by
$$C_\theta = \sigma^2 (H^T H)^{-1} \quad (10)$$
The error amplification matrix $(H^T H)^{-1}$ in (10) can be computed numerically. Since H in not orthogonal, $(H^T H)^{-1}$ is not diagonal and the θs will be correlated. The diag$(H^T H)^{-1}$ is inversely proportional to the number of samples N.

The main problem in the estimation of LFC noise is that due to the non orthogonality of H, the inverse problem quickly becomes ill-conditioned as we increase the numbers of parameters to be estimated. Figure

14 shows the condition number of $H^TH$ as a function of the number of modes in H for up to 200 modes. For large H small oscillations in the data produce large variations in the parameter vector θ.

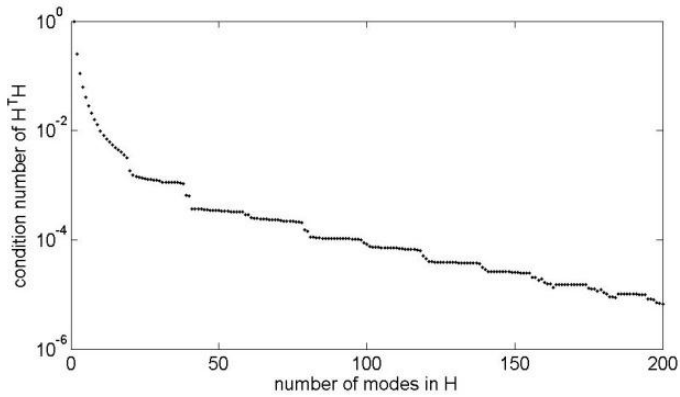

Fig. 14: Estimation error analysis. $H^TH$ condition number as a function of the number of estimated modes.

The estimator has been analyzed using SVD and generalized SVD techniques for several (H,L) pairs, where L is a linear operator. The minimization procedure was carried using Tikhonov's rule:

$$\min_\theta \{\|H\theta - y\|^2 + \|Ly\|^2\}$$

We also observe that the low frequency estimation errors have a heavier weight. Figure 15 shows the error amplification of $(H^TH)^{-1}$ normalized to N=1 for a 200 mode estimator versus a 20 mode estimator. The 200 mode estimator has almost an order of magnitude more error in the first 15 modes due to mode correlation.

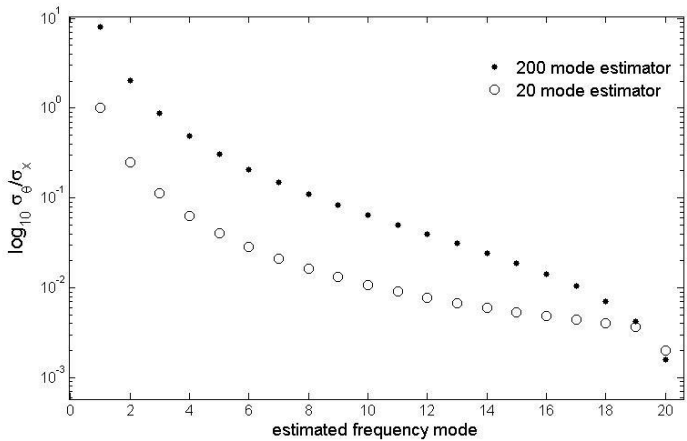

Fig. 15: Estimation error as a function of the estimated mode.

The low noise results shown in Figures 11 to 13 have been obtained using a slightly different approach. Instead of inverting a large (i.e. 200 mode matrix) a series of smaller matrices (i.e. 20 modes) are inverted and used for estimation. The reasoning is that LF noise modes that are far away from each other are uncorrelated. This method is effective and keeps the $C_\theta$ small for most $\theta_i$. However, as it can be seen in Figure 13, a few modes have a large estimation error. The optimal solution to minimize the $C_\theta$ is still open and we will keep studying the subject.

Currently, the WGN σ has been improved to 1.8nV/sqrt(Hz). The total estimation error is numerically computed and shown to be about 0.1e-.

# 7 Hardware implementation

The estimator and digital CDS reduce the CCD noise deep into the sub-electron region. The price to pay for lower noise is a more sophisticated readout system. The estimator requires an FPGA or processor capable of performing millions of multiplications per second, subtracting the estimated LFN from the original signal and digitally computing the CDS of the filtered signal. However, modern FPGAs have enough horsepower to process and filter several video channels.

The example in this paper estimates correlated noise in data sets of 20 pixels long and shows a minimum for an integration time of about 120μs. The data is sampled at 20 MHz and pre-filtered and decimated, producing a data steam of 125Ksamples/s. Each 20 pixel dataset is 5ms long and has 6000 samples. The number of multiplications required for the estimation of 2 times 200 modes is 2.4 million in 5ms or 480 million multiplications/s. The filter reconstruction requires the same number for a total of 960 million multiplications/s. Modern FPGAs have hundreds and even a couple of thousands of multipliers running up to 400 million multiplications/s. It is also worth mentioning that the pseudo inverse of H needs to be computed only once, so it can be computed off line and stored in memory. The memory requirements are N by 2M. This is 3.6MB for the current example.

Figure 16 shows a fits image taken with the FPGA based digital CDS.

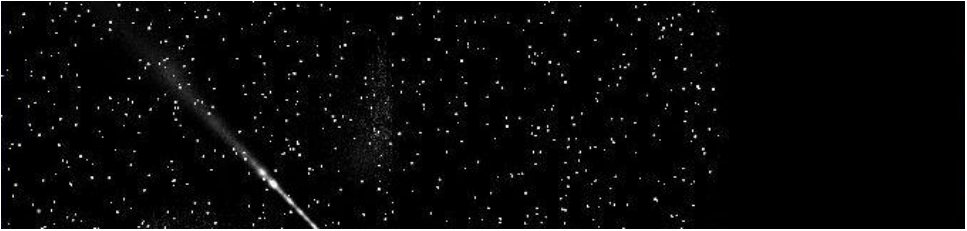

Figure 16: CCD image taken with the FPGA based digital CDS.

# 8 Conclusions and future work

Fast sampling and digital processing the CCD video signal has allowed us to reduce the low frequency noise that correlates pixel to pixel. Deep sub electron noise of $0.4e^-$ and 1-σ error bars of $0.05\ e^-$ have been systematically achieved by this method. The readout bandwidth is about 4 Kpix/s. We have also achieved $1e^-$ of noise at 50 Kpix/s by improving EMI, crosstalk, and power supply noises. Digital CCD readout systems are more sophisticated than traditional analog CDS but will allow better signal to noise ratio for spectrometers and will improve the sensitivity of CCD based detectors at very low energies. This is essential in CCD based direct Dark Matter search and neutrino experiments and also nuclear applications such as neutron imaging.

The complexity of the digital signal processing can be overcome with the use of FPGAs or embedded processors. Currently we are developing an FPGA based system that can accommodate several CCD video channels. The goal is to obtain a sub electron noise image in fits format. The search for lower noise is not finished yet. Further understanding of noise contribution in the estimator and the achievement of lower WGN will be pursued in the near future.

## Acknowledgements

We want to thank the Fermilab staff at SiDet facility, in particular Kevin Kuk, Donna Kubik, Walter Struemel who helped us with the operations of CCDs systems; and Stephen Holland from Lawrence Berkeley National Laboratory for always being available to answer our CCD questions.